# Manganese-Functionalized GelMA Hydrogels for MRI-Guided Immunotheranostics in Precision Oncology

Motahareh Nazari[1], keyvan Alavi[2*]

[1] Department of Biomedical Engineering, Science and Research Branch, Islamic Azad University, Tehran, Iran

2* Department of Biomedical Engineering, Science and Research Branch, Islamic Azad University, Tehran, Iran

Email[1]: motpotnazari@gmail.com _ motahareh.nazari0762@iau.ir

Email[2*]: keivan.alavi@gmail.com _ k.alavi@iau.ir

## Abstract

Precision oncology requires multifunctional platforms capable of integrating accurate tumor diagnosis, localized therapeutic delivery, immune modulation, and real-time monitoring of treatment response. Gelatin methacryloyl (GelMA) hydrogels have emerged as versatile biomaterials for biomedical engineering because of their biocompatibility, extracellular matrix-like structure, tunable mechanical properties, photocrosslinkability, and capacity to incorporate therapeutic agents, imaging probes, and functional nanomaterials. In parallel, manganese-based materials have gained increasing attention as promising alternatives to gadolinium-based magnetic resonance imaging contrast agents and as therapeutic components capable of modulating the tumor microenvironment. Manganese ions and manganese-based nanomaterials can enhance $T_1$-weighted MRI contrast, generate reactive oxygen species, relieve tumor hypoxia, deplete glutathione, promote immunogenic cell death, and activate the cyclic GMP–AMP synthase–Stimulator of Interferon Genes pathway. The integration of manganese-based systems with GelMA hydrogels provides a promising strategy for the development of localized, stimuli-responsive, and MRI-guided immunotheranostic platforms. This review summarizes the fundamental properties of GelMA hydrogels, the diagnostic and therapeutic roles of manganese-based materials, strategies for constructing manganese-functionalized GelMA systems, and their potential applications in precision oncology. Current challenges, including manganese-associated toxicity, controlled ion release, mechanical optimization, reproducibility, and clinical translation, are also discussed. Finally, future directions are proposed for the rational design of safe, scalable, and personalized manganese-functionalized GelMA platforms for cancer diagnosis and therapy.

**Keywords:** GelMA hydrogel; manganese; MRI contrast agent; cGAS–STING pathway; cancer immunotherapy; tumor microenvironment; theranostics; precision oncology.

# 1. Introduction

Cancer remains one of the most complex and heterogeneous diseases, characterized by uncontrolled cell proliferation, genetic and epigenetic alterations, immune evasion, metabolic reprogramming, and dynamic interactions with the tumor microenvironment. Despite major progress in surgery, chemotherapy, radiotherapy, targeted therapy, and immunotherapy, many cancers remain difficult to treat because of drug resistance, systemic toxicity, tumor recurrence, metastasis, and patient-specific biological variability(1, 2) These limitations have accelerated the development of precision oncology, a treatment paradigm that aims to tailor diagnosis and therapy according to the molecular, cellular, and immunological features of each tumor.

A major challenge in precision oncology is the need for platforms that can perform more than one function simultaneously(3, 4). Ideally, an advanced anticancer system should localize at the tumor site, provide high-resolution imaging, release therapeutic agents in a controlled manner, remodel the tumor microenvironment, activate antitumor immunity, and allow real-time monitoring of therapeutic response(5). Conventional drug delivery systems and imaging agents often perform only one of these tasks, which limits their ability to address the complexity of cancer biology.

Hydrogels have attracted considerable attention as local delivery systems and tissue-mimetic biomaterials because of their high water content, porous structure, tunable degradation, and capacity to encapsulate cells, drugs, proteins, nucleic acids, and nanoparticles(6). Among them, gelatin methacryloyl (GelMA) is one of the most widely used hydrogel platforms(7). GelMA combines the biological activity of gelatin with the photocrosslinking capacity of methacrylate groups, making it highly suitable for injectable systems, three-dimensional bioprinting, drug delivery, tumor modeling, and regenerative medicine(8, 9).

At the same time, manganese-based materials have emerged as attractive components for cancer theranostics(10). Manganese is an essential trace element with favorable paramagnetic properties, making $Mn^{2+}$ useful for $T_1$-weighted magnetic resonance imaging. Beyond imaging, manganese-based nanomaterials can respond to the tumor microenvironment, catalyze reactive oxygen species generation, relieve tumor hypoxia, deplete glutathione, and activate innate immune signaling through the cGAS–STING pathway (11-13). These properties make manganese especially valuable for the design of multifunctional platforms that combine imaging and therapy.

The integration of manganese-based systems with GelMA hydrogels offers a unique opportunity to construct localized, stimuli-responsive, and MRI-guided immunotheranostic platforms. GelMA can serve as a biocompatible matrix that retains manganese-based materials at the tumor site, controls $Mn^{2+}$ release, reduces systemic exposure, and enables combination therapy. This review focuses on the design, function, and future potential of manganese-functionalized GelMA hydrogels in precision oncology.

# 2. Fundamentals of GelMA Hydrogels

Gelatin methacryloyl is synthesized by reacting gelatin(14), a denatured derivative of collagen, with methacrylic anhydride. During this process, methacryloyl groups are introduced onto amino and hydroxyl groups of gelatin, giving the polymer photocrosslinkable functionality while preserving many of the biological motifs of gelatin. These include arginine–glycine–aspartic acid sequences that support cell adhesion and matrix metalloproteinase-sensitive sites that allow enzymatic degradation and tissue remodeling (8, 9).

Upon exposure to light in the presence of a suitable photoinitiator, GelMA undergoes polymerization to form a covalently crosslinked three-dimensional hydrogel network. This process can occur under mild aqueous conditions, making GelMA suitable for encapsulating living cells, proteins, growth factors, nucleic acids, drugs, and nanoparticles. The resulting hydrogel has a hydrated, porous, and extracellular matrix-like structure that supports molecular diffusion, cellular interaction, and localized delivery.

One of the most important advantages of GelMA is its tunability(15). Its stiffness, swelling ratio, pore structure, degradation rate, permeability, and drug-release behavior can be adjusted by changing the gelatin source, degree of methacrylation, polymer concentration, photoinitiator type, light intensity, and crosslinking time. These parameters allow GelMA hydrogels to be engineered for specific biomedical applications, including soft tissue engineering, bone regeneration, wound healing, vascular models, neural systems, drug delivery, and three-dimensional tumor models.

GelMA also possesses excellent compatibility with advanced fabrication technologies. It can be processed by three-dimensional bioprinting, microfluidics, photolithography, and injectable gelation strategies. This makes it possible to construct patient-specific structures, spatially patterned hydrogels, and complex tumor-mimetic models. In oncology, GelMA-based hydrogels can be used not only as drug delivery matrices but also as tumor models for studying cancer cell behavior, stromal interactions, immune responses, and therapeutic screening (16, 17).

However, pristine GelMA has several limitations. It may show insufficient mechanical strength, rapid degradation under some conditions, limited intrinsic bioactivity for certain applications, and dependence on photopolymerization parameters. Therefore, functionalization with nanoparticles, polymers, bioactive molecules, or inorganic components is commonly used to improve its mechanical, biological, imaging, and therapeutic properties. Manganese-based functionalization is particularly attractive because it can introduce MRI visibility, tumor microenvironment responsiveness, oxidative therapeutic activity, and immune-stimulatory effects into the GelMA network. The multifunctional mechanisms of manganese-functionalized GelMA hydrogels are illustrated in Figure 1, emphasizing their ability to combine MRI contrast enhancement, tumor microenvironment-responsive $Mn^{2+}$ release, ROS-mediated therapy, hypoxia relief, glutathione depletion, immune activation, and sustained co-delivery of therapeutic agents.

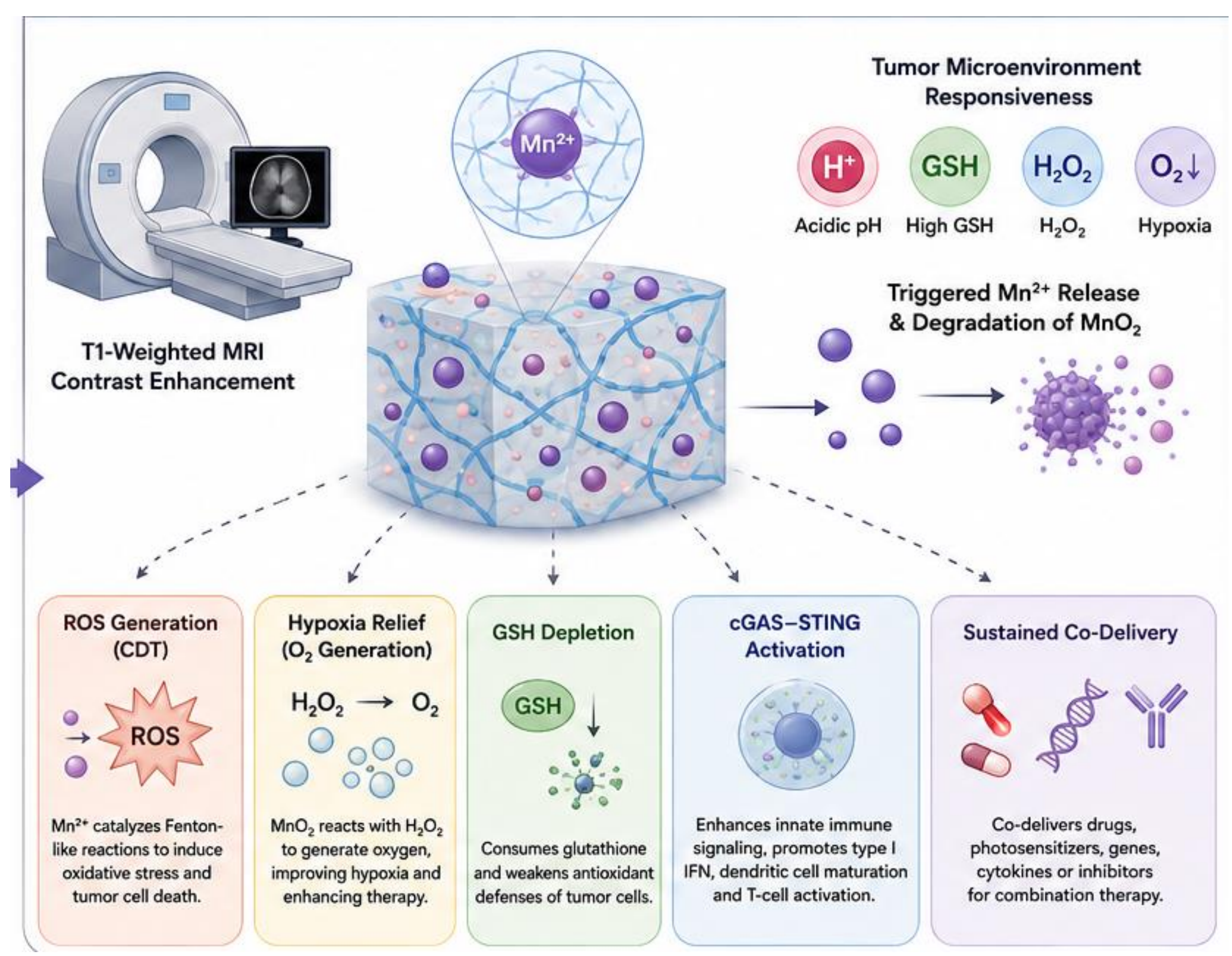

**Figure 1: Multifunctional mechanism of 1 Manganese-Functionalized GelMA Hydrogels.**

## 3. Manganese as a Diagnostic and Therapeutic Component

Manganese has gained significant attention as a functional element in cancer nanomedicine because it combines diagnostic and therapeutic properties. The divalent manganese ion has five unpaired electrons, which gives it strong paramagnetic behavior and the ability to shorten the longitudinal relaxation time of water protons. As a result, $Mn^{2+}$ can generate positive contrast in $T_1$-weighted magnetic resonance imaging (9, 13)

Manganese-based contrast agents have been explored as alternatives to gadolinium-based contrast agents. Although gadolinium agents are widely used clinically, concerns remain regarding nephrogenic systemic fibrosis in patients with renal impairment and long-term gadolinium deposition in tissues. Manganese, as an essential trace element, offers a potentially more biocompatible alternative when its release and biodistribution are properly controlled. However, free $Mn^{2+}$ is not risk-free. Excessive manganese accumulation, especially in the central nervous system, may cause neurotoxicity and manganism-like symptoms. Therefore, safe manganese-based platforms require stable coordination, controlled release, and efficient clearance.

To address these concerns, manganese has been incorporated into macrocyclic complexes, polymeric carriers, manganese oxide nanoparticles, hollow nanostructures, biomimetic nanoparticles, and tumor-responsive nanoplatforms (13, 18-23). These systems can improve stability, enhance relaxivity, reduce premature ion release, and allow tumor-specific activation. In cancer therapy, manganese-based materials are especially useful because they can respond to tumor-associated stimuli such as acidic pH, elevated glutathione, high hydrogen peroxide levels, hypoxia, and oxidative stress.

Beyond MRI, manganese-based materials can participate in several therapeutic mechanisms. $MnO_2$ nanoparticles can react with endogenous hydrogen peroxide to generate oxygen, thereby relieving tumor hypoxia and improving oxygen-dependent treatments such as photodynamic therapy and radiotherapy. Manganese-based systems can also consume intracellular glutathione, weakening tumor antioxidant defenses and increasing oxidative stress. In addition, $Mn^{2+}$ can catalyze Fenton-like reactions that generate reactive oxygen species, supporting chemodynamic therapy. Importantly, manganese can also enhance innate immune activation through the cGAS–STING pathway, making it valuable for cancer immunotherapy(11, 20, 21, 24) [5,6,15,16].

Because of this combination of MRI contrast, tumor responsiveness, oxidative therapy, oxygen generation, and immune activation, manganese is increasingly regarded as a promising building block for next-generation cancer theranostic systems. The potential precision oncology applications of manganese-functionalized GelMA hydrogels are summarized schematically in Figure 2, highlighting their roles as MRI-guided local therapeutic depots, post-surgical recurrence-prevention platforms, immune-activating systems, 3D MRI-visible tumor models, and multimodal combination-therapy scaffolds.

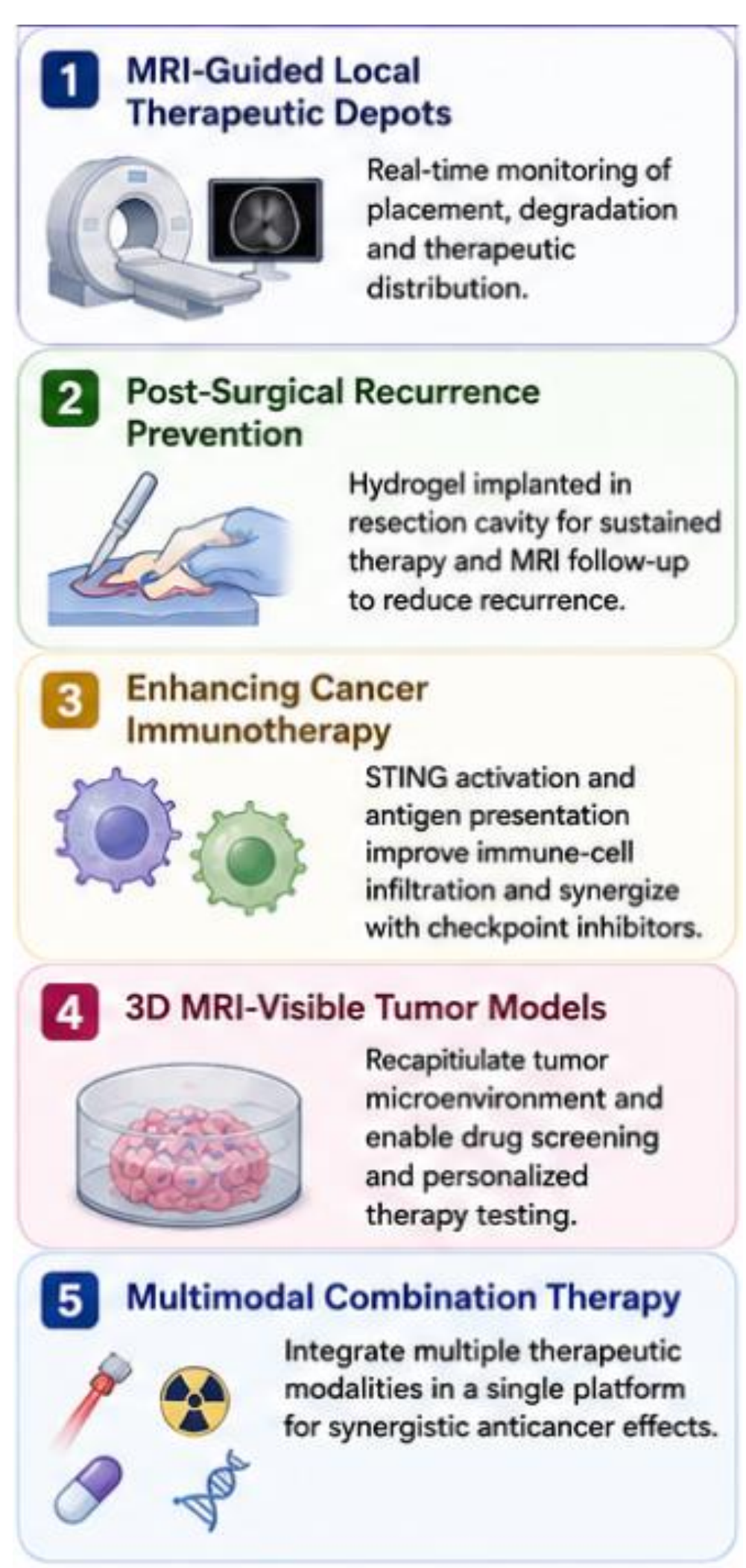

**Figure 2 Application of Manganese-Functionalized GelMA Hydrogels in Precision oncology.**

# 4. Strategies for Constructing Manganese-Functionalized GelMA Hydrogels

Manganese-functionalized GelMA hydrogels can be designed using several material strategies. The choice of strategy strongly affects imaging performance, therapeutic efficacy, release kinetics, mechanical properties, and biosafety.

One simple strategy is the physical encapsulation of manganese-based nanoparticles or manganese complexes within the GelMA precursor solution before photocrosslinking. After gelation, the manganese components become trapped within the hydrogel network. This approach is straightforward and compatible with injectable or printed systems. However, uncontrolled diffusion or burst release may occur if the interaction between manganese components and GelMA is weak.

A second approach involves incorporating manganese oxide nanoparticles, particularly $MnO_2$, into GelMA. $MnO_2$-based systems are attractive because they can respond to the acidic and reductive tumor microenvironment. Under tumor conditions, $MnO_2$ can degrade to release $Mn^{2+}$, improving MRI contrast while simultaneously consuming glutathione and generating oxygen from hydrogen peroxide. This makes $MnO_2$-GelMA systems suitable for MRI-guided photodynamic therapy, chemodynamic therapy, and immunotherapy.

A third strategy is to use manganese chelates or coordination complexes within the hydrogel network. Stable chelation can reduce premature $Mn^{2+}$ release and improve safety. Chelated manganese systems may be physically loaded into GelMA or chemically modified to interact more strongly with the hydrogel matrix. This strategy is useful when the primary goal is safe and sustained MRI contrast.

A fourth strategy involves hybrid nanoplatforms, where manganese is combined with other therapeutic or imaging components. For example, manganese-based nanoparticles may be integrated with photosensitizers, chemotherapeutic agents, immune adjuvants, photothermal agents, or targeting ligands before encapsulation into GelMA. Such systems can support multimodal imaging and combination therapy.

A fifth strategy is covalent or affinity-based functionalization, where manganese-binding ligands or manganese-containing structures are chemically linked to GelMA. This can improve retention and reduce burst release, but it requires more complex synthesis and careful control of hydrogel chemistry.

Overall, the ideal manganese-functionalized GelMA system should provide stable incorporation before administration, tumor-specific activation after delivery, controlled $Mn^{2+}$ release, strong MRI contrast, localized therapeutic activity, and minimal systemic toxicity. The design and local delivery strategy of manganese-functionalized GelMA hydrogels is illustrated in Figure 3, showing the incorporation of manganese-based nanomaterials or $Mn^{2+}$ ions into the GelMA precursor to form an injectable or implantable hydrogel platform for localized tumor treatment.

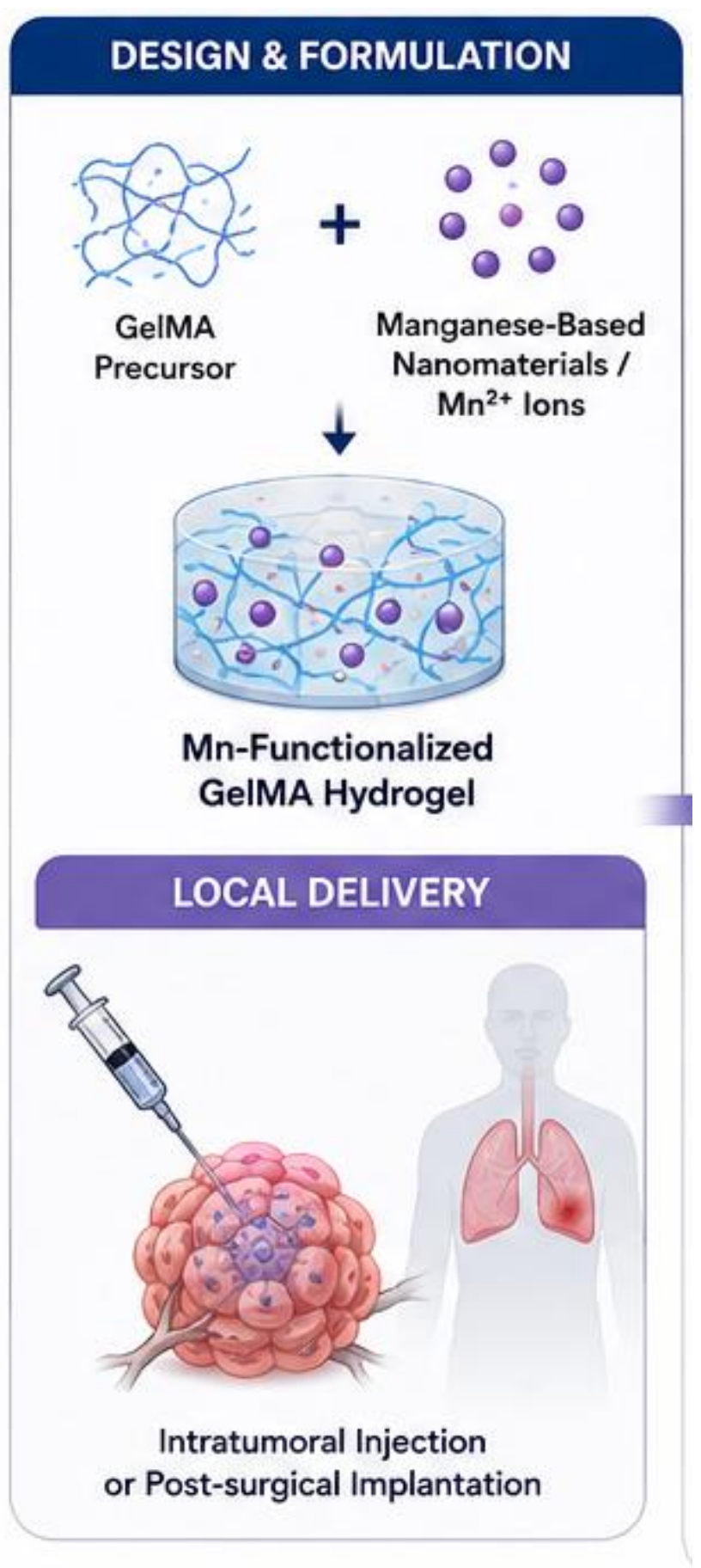

**Figure 3 Design, formulation, and local delivery of manganese-functionalized GelMA hydrogels**

## 5. MRI-Guided Function of Manganese-Functionalized GelMA Hydrogels

Magnetic resonance imaging is one of the most powerful non-invasive imaging techniques in oncology because it provides high spatial resolution and excellent soft-tissue contrast. However, many tumors require contrast agents to improve detection, define tumor margins, monitor therapeutic distribution, and evaluate treatment response. Manganese-functionalized GelMA hydrogels can serve as localized MRI-visible depots for cancer diagnosis and therapy monitoring.

In these systems, manganese provides $T_1$-weighted MRI contrast, while GelMA provides spatial localization and controlled release. When implanted or injected near a tumor site, the hydrogel can retain manganese-based components locally, reducing systemic distribution. Tumor microenvironment-responsive manganese release can further enhance imaging selectivity. For example, acidic pH, high glutathione levels, or elevated hydrogen peroxide may trigger $MnO_2$ degradation and $Mn^{2+}$ release, producing stronger MRI signals in tumor tissue than in normal tissue.

MRI-visible GelMA hydrogels offer several advantages for precision oncology. First, they can help confirm accurate hydrogel placement after injection or implantation. Second, they can allow non-invasive monitoring of hydrogel degradation and therapeutic release. Third, they can track the distribution of manganese-based nanomaterials or co-loaded drugs. Fourth, they may allow real-time evaluation of treatment response when combined with therapeutic modalities such as chemotherapy, photodynamic therapy, photothermal therapy, chemodynamic therapy, or immunotherapy.

The performance of manganese-functionalized GelMA MRI systems depends on several parameters, including manganese concentration, relaxivity, hydrogel pore size, swelling behavior, degradation rate, ion-release profile, and interaction between water molecules and manganese centers. Excessively strong retention may reduce MRI sensitivity, whereas rapid release may increase toxicity. Therefore, rational tuning of hydrogel structure and manganese chemistry is essential.

## 6. Therapeutic Functions Beyond Imaging

Although manganese is valuable as an MRI contrast component, its therapeutic functions are equally important. Manganese-based systems can directly participate in tumor treatment through oxidative stress regulation, hypoxia relief, drug delivery, and immune activation.

One major therapeutic mechanism is chemodynamic therapy. In the tumor microenvironment, released $Mn^{2+}$ can catalyze Fenton-like reactions that generate reactive oxygen species. These reactive oxygen species damage proteins, lipids, and DNA, leading to tumor cell death. When manganese-based catalysts are retained within GelMA hydrogels, ROS generation can be localized to the tumor region, potentially reducing off-target toxicity.

Another important mechanism is hypoxia modulation. Tumor hypoxia reduces the efficacy of radiotherapy, photodynamic therapy, chemotherapy, and immunotherapy. $MnO_2$-based nanoparticles can react with endogenous hydrogen peroxide to produce oxygen. When incorporated into GelMA, these nanoparticles can act as local oxygen-generating reservoirs, improving the tumor microenvironment and enhancing oxygen-dependent therapies.

Manganese-based systems can also deplete glutathione, a major intracellular antioxidant that protects tumor cells from oxidative damage. By reducing glutathione levels, $MnO_2$ and related materials can weaken tumor antioxidant defenses and amplify ROS-mediated therapy. This effect is particularly useful when combined with photodynamic therapy, chemodynamic therapy, or radiotherapy.

In addition, GelMA hydrogels can co-deliver manganese-based materials with chemotherapeutic agents, photosensitizers, immune adjuvants, checkpoint inhibitors, proteins, nucleic acids, or natural bioactive compounds. The hydrogel matrix provides sustained release and local retention, while manganese contributes to imaging visibility and therapeutic enhancement.

# 7. Manganese-Mediated cGAS–STING Activation and Cancer Immunotherapy

The cGAS–STING pathway is a central innate immune signaling axis involved in the detection of cytosolic DNA. Activation of this pathway promotes the production of type I interferons and inflammatory cytokines, stimulates dendritic cell maturation, enhances antigen presentation, and supports cytotoxic T-cell and natural killer cell responses. Because many tumors suppress immune activation, strategies that stimulate cGAS–STING signaling are highly attractive for cancer immunotherapy (2, 24, 25)

Manganese has been shown to potentiate cGAS–STING signaling. $Mn^{2+}$ can increase the sensitivity of cGAS to cytosolic DNA and enhance downstream STING activation. In cancer therapy, this can promote innate immune activation and contribute to the conversion of immunologically “cold” tumors into more inflamed “hot” tumors. Such immune remodeling may improve responsiveness to immune checkpoint blockade therapies.

Manganese-functionalized GelMA hydrogels are especially promising for STING-related immunotherapy because they can provide localized and sustained manganese delivery. Local hydrogel-based delivery may reduce systemic inflammatory toxicity while maintaining effective immune stimulation at the tumor site. This is particularly relevant for post-surgical cancer therapy, where an injectable or implantable hydrogel could be placed in the tumor resection cavity to prevent recurrence by combining MRI monitoring, local therapy, and immune activation.

However, STING activation requires careful control. Excessive, prolonged, or systemic activation may cause inflammatory damage and immune-related adverse effects. Therefore, future Mn-GelMA immunotherapy systems should be designed to regulate the amount, location, and duration of $Mn^{2+}$ release. Important evaluation markers should include type I interferon production, dendritic cell maturation, macrophage polarization, $CD8^{+}$ T-cell infiltration, cytokine profiles, immune memory formation, and synergy with checkpoint inhibitors.

## 8. Tumor Microenvironment-Responsive Mn-GelMA Platforms

The tumor microenvironment contains several abnormal features that can be exploited for selective therapy. These include acidic pH, high glutathione concentration, elevated hydrogen peroxide, hypoxia, abnormal enzyme expression, inflammation, and oxidative stress. Smart manganese-functionalized GelMA hydrogels can be engineered to respond to these features and release therapeutic agents only under tumor-associated conditions.

pH-responsive systems are useful because many tumors are more acidic than normal tissues. Acidic conditions can accelerate the degradation of certain manganese-based nanoparticles or acid-labile linkages, promoting $Mn^{2+}$ release and drug delivery. Redox-responsive systems are also attractive because tumors often contain elevated glutathione levels. $MnO_2$ can react with glutathione, leading to $Mn^{2+}$ generation, glutathione depletion, and enhanced oxidative stress.

ROS-responsive GelMA systems can be designed using oxidation-sensitive linkages such as boronate bonds or thioketal groups. These systems may release antioxidants, antibiotics, anticancer drugs, or immune modulators in response to oxidative stress. Although many ROS-responsive GelMA examples have been developed for bone regeneration and infection treatment, similar principles can be adapted to cancer therapy by designing hydrogels that respond to tumor oxidative imbalance.

Enzyme-responsive systems may use matrix metalloproteinase-sensitive sequences or tumor-associated enzymatic degradation to control hydrogel breakdown and cargo release. Because GelMA naturally contains enzymatically degradable motifs, its degradation behavior can be tuned for tumor-specific release.

Overall, tumor-responsive Mn-GelMA hydrogels are promising because they can improve selectivity, reduce systemic toxicity, and synchronize imaging with therapeutic activation.

## Table 1. Functional roles of manganese-functionalized GelMA hydrogels in precision oncology

| Functional element | Main mechanism | Role in cancer diagnosis/therapy | Advantage of GelMA integration | Key limitation | Ref. |
|---|---|---|---|---|---|
| $Mn^{2+}$ ions or manganese complexes | Shorten $T_1$ relaxation time of water protons | Positive MRI contrast and treatment monitoring | Local depot effect and controlled release | Risk of uncontrolled $Mn^{2+}$ release and neurotoxicity | (26) |
| $MnO_2$ nanoparticles | TME-responsive degradation, GSH depletion, oxygen generation | MRI enhancement, hypoxia relief, PDT/CDT/PTT support | Localized retention and sustained activity | Release kinetics and degradation must be controlled | (27) |
| Hollow manganese nanostructures | High surface area and internal drug-loading space | MRI-guided drug delivery and combination therapy | Improved local loading and retention | Complex synthesis and reproducibility issues | (22) |
| Mn-mediated ROS generation | Fenton-like catalytic reactions | Chemodynamic therapy and oxidative tumor damage | Local ROS production near tumor site | Depends on tumor pH and $H_2O_2$ concentration | (28) |
| Mn-enhanced cGAS–STING activation | Amplifies innate immune signaling | Antitumor immunity and checkpoint blockade sensitization | Local immune stimulation with reduced systemic exposure | Risk of excessive inflammation | (29) |
| Multimodal imaging systems | Combination with fluorescence, PAI, CT, PET, or photothermal agents | Biodistribution tracking and response monitoring | Co-loading of imaging and therapeutic agents | Increased regulatory and safety complexity | (30) |
| Smart GelMA network | pH, ROS, enzyme, redox, light, or magnetic responsiveness | Controlled release and tumor-specific activation | Tunable degradation, swelling, and mechanics | Requires careful optimization for each application | (31) |

# 9. Applications in Precision Oncology

Manganese-functionalized GelMA hydrogels have several plausible applications in precision oncology, but their translational maturity should be described cautiously. The most clinically credible near-term use-case is as a localized postoperative or intratumoral depot for tumors accessible to direct administration, where the hydrogel could retain therapeutic payloads at the disease site while manganese provides MRI visibility for placement verification and longitudinal monitoring. However, MRI visibility alone is not a clinical endpoint; future studies must show that such systems improve decision-making, reduce systemic exposure, or enhance local control compared with current perioperative imaging and treatment workflows.

The strongest precision-oncology rationale for Mn-GelMA systems lies in prevention of post-surgical recurrence. Human studies with personalized cancer vaccines have already shown that the postoperative setting can support durable immune control in high-risk cancers, including kidney cancer, pancreatic cancer, and melanoma. These findings suggest that Mn-GelMA platforms should be framed not simply as local drug reservoirs, but as candidate minimal-residual-disease interventions that may complement perioperative immunotherapy. In this context, the critical benchmarks are recurrence-free survival, locoregional control, and durable systemic immune activation rather than short-term tumor shrinkage alone.

Mn-GelMA systems may also enhance immunotherapy through local manganese-mediated amplification of cGAS–STING signaling, but this remains a mechanistically attractive hypothesis rather than a clinically validated outcome. A more realistic translational position is to develop these hydrogels as adjuncts to established perioperative immunotherapies, including checkpoint blockade, especially in settings where local immune priming and postoperative recurrence control are both needed. Local immune therapies such as talimogene laherparepvec provide proof that site-directed immunologic intervention can have durable antitumor effects in patients, but Mn-GelMA platforms still need to demonstrate comparable durability and safety.

In three-dimensional tumor modeling, GelMA remains highly relevant because it can reproduce extracellular-matrix-like features and support ex vivo drug testing, while manganese functionalization may enable longitudinal, non-destructive imaging of material degradation or local physicochemical change. Yet the clinical value of these models will depend on whether they improve therapy selection or response prediction, not simply whether they are MRI-visible. Similarly, combination-therapy applications are promising, but future work should prioritize clinically coherent pairings—especially postoperative immune intensification and image-guided local-plus-systemic treatment—over increasingly complex multifunctional constructs with uncertain manufacturing and regulatory paths(32).

## 10. Current Challenges

Despite their promise, manganese-functionalized GelMA hydrogels face several important challenges before clinical translation.

The first challenge is manganese safety(10). Although manganese is an essential element, excessive $Mn^{2+}$ exposure can cause toxicity, especially in the nervous system. Future systems must carefully control manganese dose, release rate, biodistribution, metabolism, and clearance. Long-term toxicity studies are essential.

The second challenge is controlled release(10). A successful platform must release enough $Mn^{2+}$ to generate MRI contrast and therapeutic effects, but not so much that systemic toxicity occurs. This requires precise tuning of hydrogel crosslinking density, pore size, degradation rate, nanoparticle composition, and manganese coordination chemistry.

The third challenge is mechanical and structural optimization(33). GelMA hydrogels can be mechanically weak, especially under physiological conditions. Reinforcement strategies such as double-network hydrogels, nanocomposite design, hybrid polymers, and optimized photocrosslinking may be needed.

The fourth challenge is reproducibility. Variations in GelMA synthesis, degree of methacrylation, nanoparticle size, manganese loading, and photocrosslinking conditions can strongly affect performance. Standardized synthesis and characterization protocols are required.

The fifth challenge is immunological control. Manganese-mediated STING activation is promising but must be carefully regulated. Excessive activation may cause inflammatory toxicity, while insufficient activation may fail to produce meaningful antitumor immunity.

The sixth challenge is clinical translation. Multifunctional platforms are scientifically attractive but may face regulatory difficulties because they combine biomaterials, imaging agents, nanomedicine, and therapeutics. Scalable synthesis, sterilization, storage stability, good manufacturing practice production, and regulatory classification must be considered early in development.

## Table 2. Current challenges and future directions for manganese-functionalized GelMA hydrogels

| Challenge | Why it matters | Suggested future direction |
|---|---|---|
| $Mn^{2+}$ toxicity and tissue accumulation | Uncontrolled release may cause neurotoxicity and long-term safety concerns | Develop stable chelates, biodegradable Mn systems, localized delivery, and detailed biodistribution studies |
| Limited direct Mn-GelMA oncology studies | Many reports discuss GelMA or manganese separately | Generate more in vivo cancer studies using integrated Mn-GelMA platforms |
| Burst release from hydrogels | Rapid release may reduce efficacy and increase toxicity | Use covalent binding, affinity interactions, nanoparticle encapsulation, or multilayer hydrogel design |
| Weak mechanical strength of GelMA | May limit implantation, injection stability, or long-term retention | Use double-network hydrogels, reinforced composites, or hybrid polymer systems |
| Insufficient quantitative comparison | Difficult to compare platforms without standard metrics | Report relaxivity, Mn release, modulus, degradation, swelling, drug loading, tumor inhibition, and immune markers |
| STING overactivation | Excessive immune stimulation may cause inflammation | Control local $Mn^{2+}$ dose and monitor cytokines, immune-cell infiltration, and systemic toxicity |
| Complex multifunctional designs | Highly complex systems may be hard to manufacture or regulate | Prioritize modular, scalable, and clinically realistic designs |
| Lack of standardization | Batch variability limits reproducibility | Standardize GelMA synthesis, Mn loading, nanoparticle characterization, and biological testing |
| Limited long-term in vivo data | Short-term studies may miss delayed toxicity | Conduct pharmacokinetic, biodistribution, clearance, and chronic toxicity studies |
| Regulatory uncertainty | Combination products face difficult approval pathways | Design with GMP production, sterilization, storage, and clinical workflow in mind |

# 11. Future Perspectives

Future progress in manganese-functionalized GelMA hydrogels will depend on rational material design, biological validation, and translational planning.

One important direction is the development of safer manganese coordination systems. Strong chelators, biodegradable manganese nanostructures, and tumor-triggered release mechanisms may reduce systemic toxicity while maintaining imaging and therapeutic efficacy.

Another direction is AI-assisted hydrogel design. Machine learning models could be used to predict GelMA stiffness, printability, degradation, swelling, drug release, and biological performance based on formulation parameters. This may accelerate optimization and reduce experimental trial-and-error.

Three-dimensional bioprinting is also expected to play an important role. Mn-functionalized GelMA bioinks could be used to fabricate patient-specific tumor models or implantable therapeutic scaffolds. Such systems may support personalized drug screening, immune-response evaluation, and MRI-visible tissue constructs.

Combination immunotherapy is another promising area. Mn-GelMA hydrogels could be combined with checkpoint inhibitors, cancer vaccines, dendritic cell activators, or adoptive cell therapy. Local hydrogel delivery may improve immune activation at the tumor site while reducing systemic immune-related adverse effects.

Finally, clinical translation will require simpler and more scalable designs. Although multifunctional systems are attractive, overly complex platforms may be difficult to manufacture and approve. Future studies should prioritize clear therapeutic mechanisms, reproducible synthesis, scalable production, and clinically relevant administration routes.

# 12. Conclusion

Manganese-functionalized GelMA hydrogels represent a promising class of multifunctional biomaterials for MRI-guided cancer theranostics. GelMA provides a biocompatible, tunable, and localizable hydrogel matrix, while manganese-based components contribute MRI contrast, tumor microenvironment responsiveness, ROS generation, hypoxia relief, glutathione depletion, and cGAS–STING-mediated immune activation. Together, these features make Mn-GelMA systems attractive for precision oncology, particularly for localized therapy, post-surgical recurrence prevention, image-guided treatment, and immunotherapy enhancement.

However, the field remains largely preclinical, and several challenges must be addressed before clinical application. These include controlling $Mn^{2+}$ release, preventing long-term toxicity, improving hydrogel mechanical performance, standardizing synthesis, validating immune effects, and demonstrating safety in long-term in vivo studies. With continued advances in biomaterials engineering, nanomedicine, MRI technology, immunotherapy, and AI-guided hydrogel design, manganese-functionalized GelMA hydrogels may become powerful platforms for next-generation precision cancer diagnosis and therapy.

## 13. Declaration

The schematic figures were prepared by the authors using digital design tools and schematic icon-based illustration. The scientific content, conceptual organization, labels, and final interpretation were designed, checked, and approved by the authors. No published figure, microscopy image, experimental image, or copyrighted scientific illustration was reproduced or adapted.